\documentclass[prl,notitlepage,groupedaddress,twocolumn,showpacs]{revtex4-1}
\usepackage{amsmath,amssymb,graphicx,fancyhdr,subfigure,hyperref,multirow,tikz}

\newcommand{\dd}[1]{\mathrm{d}#1\,}
\newcommand{\avg}[1]{\langle{#1}\rangle}
\renewcommand{\Re}{\mathop{\mathrm{Re}}}
\renewcommand{\Im}{\mathop{\mathrm{Im}}}

\newcommand{\sgn}{\mathop{\mathrm{sgn}}}

\newcommand{\fixme}[1]{\begingroup\color{red}\em(FIXME: #1)\endgroup}

\renewcommand{\fixme}[1]{}

\begin{document}

\title{Dephasing of spin and charge interference in helical Luttinger liquids}
\author{Pauli Virtanen}
\affiliation{
  Institute for Theoretical Physics and Astrophysics, University of W\"urzburg,
  D-97074 W\"urzburg, Germany.
}
\author{Patrik Recher}
\affiliation{
  Institute for Theoretical Physics and Astrophysics, University of W\"urzburg,
  D-97074 W\"urzburg, Germany.
}
\date{\today}

\pacs{73.23.-b, 71.10.Pm}

\begin{abstract}
  We consider a four-terminal Aharonov-Bohm interference setup formed
  out of two edges of a quantum spin Hall insulator, supporting
  helical Luttinger liquids (HLLs). We show that the temperature and
  bias dependence of the interference oscillations are linked to the
  amount of spin flips in tunneling between two HLLs which is a unique
  signature of a HLL.  We predict that spin dephasing depends on the
  electron-electron ($e$-$e$) interaction but differently from the
  charge dephasing due to distinct dominant tunneling excitations. In
  contrast, in a spinful Luttinger liquid with SU(2) invariance, uncharged
  spin excitations can carry spin current without dephasing in spite
  of the presence of $e$-$e$ interactions.
\end{abstract}

\maketitle


Quantum spin-Hall (QSH) insulators can support edge states that are
topologically protected against disorder, and are present in the
absence of a magnetic field \cite{kane2005-qsh, bernevig2006-qsh,
  wu2006-hla, xu2006-soq}. These edge states have a special helical
structure, in that their direction of propagation is associated with a
given value of spin polarization. Evidence of the existence of such
helical edge states has been recently found in HgTe based quantum well
structures \cite{konig2007-qsh,roth2009-nti}. They in principle enable
direct control of spin currents via electronic means, which has
generated interest in their spintronics applications
\cite{maciejko2009-sae, jiang2009-qsh}.

However, the fact that spin can be controlled electronically implies
that it must also couple to the electron-electron ($e$-$e$)
interaction. This is in contrast to SU(2) symmetric systems in which
propagating spin excitations can be uncharged, and therefore remain
unaffected by Coulomb forces \cite{giamarchi2004-qpi}. The $e$-$e$
interaction is the dominant cause for dephasing of electronic
coherence in low-dimensional mesoscopic systems, at temperatures low
enough \cite{datta1999-eti}, and in 1D systems, this dephasing is
associated with fractionalization of excitations
\cite{lehur2002-efd}.  In QSH systems, one would therefore expect that
dephasing originating from the Coulomb interaction would be visible
also in the spin coherence, and lead to fractionalization of spin
\cite{das2010-sps}. On the one hand, this will limit the spin
coherence time which may be harmful for some applications, but on the
other hand, it can serve as a characteristic signature of the QSH
state.

One possible way to probe the coherence in a low-dimensional
mesoscopic system is to observe the decay of interference oscillations
\cite{lehur2005-dom,*lehur2006-eli}. A simple geometry in which this
can be done is an Aharonov-Bohm (AB) interferometer, where the phase
difference between two paths of propagation is controlled with a
magnetic flux. Although such flux would not ordinarily couple to spin
currents, the structure of helical edge states makes this possible
\cite{maciejko2009-sae}.  Alternatively, one can observe the
interference oscillations as a function of gate or bias voltages, or
spin density differences, applied to the system --- these will be
present also in non-helical edge states.

\begin{figure}
 \includegraphics{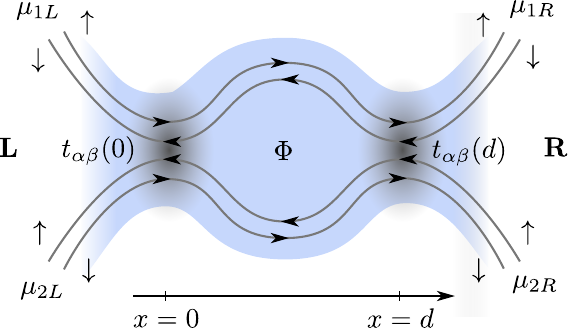}
  \caption{
    \label{fig:interferometer}
    (Color online)
    Interferometer supporting helical edge states, where the direction of
    propagation is correlated with the spin
    ($\uparrow/\downarrow$).  The four-terminal setup is connected to
    noninteracting leads, biased at potentials $\mu_{1/2,L/R}$.
    Tunneling between the edges occurs at points $x=0$ and $x=d$ (shaded).
    The loop between the tunneling points is threaded by a magnetic
    flux $\Phi$.
  }
\end{figure}

Here, we analyze dephasing of charge and spin interference
in an AB interferometer composed of tunnel-coupled interacting helical
edge states (see Fig.~\ref{fig:interferometer}).  We find that
dephasing in charge tunneling currents depends on the degree of
spin-flips in tunneling.  Moreover, we establish a direct relationship
between charge and spin currents in helical four-probe setups, which
translates the results on charge dephasing to apply to spin. Here,
helical edge states turn out to differ significantly from ordinary
spinful edge states, which in the presence of SU(2) spin symmetry can
support interference effects in spin current without dephasing, even
when Coulomb interaction is present.

{\it Model.}  An interacting helical edge state can be modeled as a
helical Luttinger liquid (HLL) \cite{wu2006-hla}. Each HLL has the
same degrees of freedom as a spinless LL (SLLL) with right ($+$) and
left ($-$) movers associated with a given spin state
($\lvert+\rangle\,\widehat{=}\lvert\uparrow\rangle$,$\lvert-\rangle\,\widehat{=}\lvert\downarrow\rangle$),
so spin is redundant. Microscopically, the spin states
$\lvert\uparrow\rangle$ and $\lvert\downarrow\rangle$ refer to Kramers
partners which are either electron spin or total angular momentum
eigenstates, depending on the material.

The bosonized Hamiltonian for a HLL reads
\cite{wu2006-hla}
\begin{equation}
  \label{eq:bosonized-hamiltonian}
    H_0 = \frac{v_F}{2}\int_{-\infty}^\infty\dd{x}
    [g(x)^{-2}:(\partial_x\vartheta)^2:+ :(\partial_x\phi)^2:]
\end{equation}
where the standard boson fields $\vartheta(x)$, $\phi(x)$ satisfy
$[\phi(x),\vartheta(x')]=(i/2)\sgn(x-x')$ and are associated with
annihilation operators for left- and right-going electrons,
$\psi_\alpha(x)\propto e^{i\alpha[k_Fx
  +\sqrt{\pi}\vartheta(x)]+i\sqrt{\pi}\phi(x)}$, $\alpha=\pm$, and
$g(x)$ is an $e$-$e$ interaction parameter. Moreover, $v_F$, $k_{F}$
are the Fermi velocity and Fermi wave vector.  For a pair of
edge states, the spin-direction mapping is reversed on the second
edge. We assume the edge states are contacted to noninteracting Fermi
leads a distance $l$ away from the central region. That is, there is a
repulsive interaction ($g(x)=g<1$) in the central region ($-l<x<l$),
but the leads ($|x|>l$) are noninteracting ($g=1$).  Here, we assume
that the relevant lead modes coupled to the system are also described
by Eq.~\eqref{eq:bosonized-hamiltonian}.  We will use dimensionless
units in which $\hbar=e=k_B=1$.

The tunneling between the upper and lower edges is described by the
Hamiltonian
\begin{align}
  H_T = L \sum_{R=0,d} \sum_{\alpha\beta=\pm} t_{\alpha\beta}(R)
  \psi_{1,\alpha}(R)^\dagger\psi_{2,\beta}(R)
  + \text{h.c.}
  \,,
\end{align}
where $j=1,2$ denote the upper and lower edges and $L$ is the total
length of an edge (including the leads).  Under time-reversal symmetry
broken only by the magnetic flux $\Phi$ in the interferometer loop,
the tunneling elements read $t_{\alpha\beta}(R) = e^{-2\pi i (R/d)
  \Phi/\Phi_0}t'_{\alpha\beta}(R)$, $t'_{\alpha\beta} = -\alpha\beta
(t_{-\alpha,-\beta}')^*$, where $\Phi_0=h/e$ is the magnetic flux
quantum.  In the presence of inversion symmetry, spin is conserved
\cite{bernevig2006-qsh} and, as a consequence, the spin-flipping terms
should vanish, $t'_{++}=t_{--}'\simeq 0$. However, local gates
\cite{vayrynen2010-ema} or strain can induce spin-mixing in the
tunneling amplitudes, allowing for non-zero $t'_{\alpha\alpha}$.

We compute currents in the presence of tunneling, treating $H_T$ as a
perturbation. This is expected to be valid in HLLs, as at low energies
the tunneling stays irrelevant (in the renormalization group sense)
for $1/2<g<2$ \cite{teo2009-cbo,hou2009-cja}.  Bias voltages in
terminals are taken into account by assuming that the incoming states
have thermal populations described by chemical potentials
$\mu_{1/2,\pm}\equiv\mu_{1/2,L/R}$ and a temperature $T$. These potentials
can be gauged into the tunneling Hamiltonian
\cite{peca2003-fia,dolcini2005-tpo}:
$t_{\alpha\beta}(R)\mapsto{}t_{\alpha\beta}(R)e^{i(\mu_{1\alpha}-\mu_{2\beta})t}$
in the interaction picture. That the system is contacted to Fermi
leads implies that despite any fractionalization to counterpropagating
plasmons, all charge injected to the $+$ ($-$) channel finally enters
the right (left) lead \cite{safi1995-tts}.  Moreover, we assume that
$l$ is large compared to the interferometer size $d$ and length scales
$\hbar/T$, $\hbar/V$ given by the bias and temperature, so that we can
neglect any finite-$l$ effects \cite{dolcini2005-tpo} on the tunneling
dynamics.

%
Within this approach, the total tunneling current from the upper edge
to the lower edge is found by computing $\avg{\hat{I}_{\rm T}}$ via the
Kubo approach (cf. \cite{lehur2005-dom,*lehur2006-eli,geller1997-aei}).
The tunneling current operator can be identified as
$\hat{I}_{\rm T}=\hat{I}_++\hat{I}_-$, where
\begin{align}
  \hat{I}_\alpha = iL \sum_{R=0,d} \sum_{\beta=\pm} t_{\alpha\beta}(R)
  \psi_{1,\alpha}(R)^\dagger\psi_{2,\beta}(R)
  + \text{h.c.}
\end{align}
describe tunneling into the $+$ and $-$ channels.

{\it Charge tunneling current.}  The leading contribution to the
charge tunneling current is given by single-particle processes, as
described by $H_T$, in the range $1/2<g<\sqrt{3}$
\cite{hou2009-cja,teo2009-cbo}. In the loop geometry, there are three
distinct ones: direct tunneling current through each contact
separately ($|t_{\alpha\beta}(x)|^2$), an interference contribution
involving both contacts
($t_{\alpha,-\alpha}(0)t_{\alpha,-\alpha}(d)^*$), and an interference
contribution with spin flips in tunneling
($t_{\alpha,\alpha}(0)t_{\alpha,\alpha}(d)^*$), the latter two in
general coupling to the AB phase. In the leading order in tunneling,
particle conservation forbids mixed processes (e.g. $t_{++}t_{+-}$).

For later convenience, we first consider the current tunneling into the
$\alpha=\pm$ channels separately, $\avg{I_{\alpha}}=
I^{\rm direct}_{\alpha} + I^{\rm AB}_{\alpha} + I^{\rm AB, sf}_{\alpha}$.
These are given by the expressions
\begin{subequations}
\label{eq:channel-current-components}
\begin{align}
  I^{\rm direct}_\alpha
  &=
  \sum_{R=0,d}
  \sum_{\beta=\pm}
  |t_{\alpha\beta}(R)|^2
  \Re Z_1(0, \mu_{1,\alpha} - \mu_{2,\beta})
  \\
  I^{\rm AB}_\alpha
  &=
  2\Re[t_{\alpha,-\alpha}(0)t_{\alpha,-\alpha}(d)^* e^{-2i\alpha{}k_Fd}]
  \\\notag&\qquad\times
  \Re Z_2(d, \mu_{1,\alpha} - \mu_{2,-\alpha})
  \,,
  \\
  I^{\rm AB, sf}_\alpha
  &=
  2\Re[
  t_{\alpha\alpha}(0)t_{\alpha\alpha}(d)^* Z_{1,\alpha}(d, \mu_{1,\alpha} - \mu_{2,\alpha})]
 \end{align}
\end{subequations}
where $Z_{1,+}=Z_{1,-}^*=Z_1$, with
$
  Z_1(x,\omega) =
  iL^2
  \int_{-\infty}^\infty\dd{t}e^{i\omega t}\bigl\langle
  [(\psi_{1,+}^\dagger\psi_{2,+})(x,t),
  (\psi_{2,+}^\dagger\psi_{1,+})(0,0)]
  \bigr\rangle_0
$
and
$
Z_2(x,\omega) =
iL^2e^{2ik_Fx}
\int_{-\infty}^\infty\dd{t}e^{i\omega t}\bigl\langle
[(\psi_{1,+}^\dagger\psi_{2,-})(x,t),\\
(\psi_{2,-}^\dagger\psi_{1,+})(0,0)]
\bigr\rangle_0.
$
These equilibrium correlators of the clean system can be found via
bosonization techniques \cite{giamarchi2004-qpi}, and the time integrals
can be evaluated analytically.

\begin{figure}
  \includegraphics{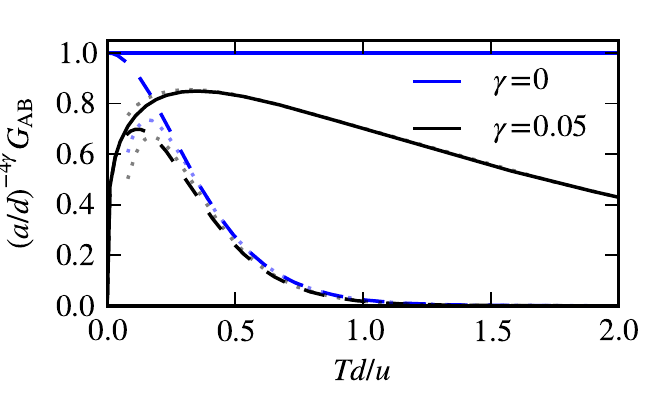}
  \caption{
    \label{fig:GAB-vs-T}
    (Color online)
    Temperature dependence of the amplitude of the dimensionless
    conductance corresponding to interference oscillation amplitudes,
    $G = (2\pi n_F^2)^{-1}\Re\,\dd{Z_{1/2}}/\dd{V}\rvert_{V=0}$.  The
    spin-flip $G_{\rm AB,sf}$ (solid), non-spin-flip $G_{\rm AB}$
    (dashed), and the high-temperature scaling from
    Eq.~\eqref{eq:high-t-dephasing} (dotted) are shown.
  }
\end{figure}

We concentrate on the tunneling charge current in a situation
where a bias $V$ is applied between the upper and lower edges,
$\mu_{1,\alpha}=V/2$, $\mu_{2,\alpha}=-V/2$.  At high temperatures
($\pi T/u\equiv{}z\gg{}1/d$), we observe that the AB oscillations
with flux $\Phi$ experience exponential dephasing (see
Fig.~\ref{fig:GAB-vs-T}), which for $V\ll T$ gives
\begin{subequations}
\label{eq:high-t-dephasing}
\begin{align}
  \label{eq:high-t-dephasing-z2}
  I_{\rm AB}
  &\propto
  Z_2
  \simeq
  \frac{
    4^{2\gamma+1}2\pi n_F^2v_F^2 z(az)^{4\gamma}
    \sin\frac{d V}{u}
  }{
    u
  }
  e^{-4(\gamma+1/2)zd}
  \,,
  \\
  \label{eq:high-t-dephasing-z1}
  I_{\rm AB,sf}
  &\propto
  Z_1
  \simeq
  \frac{
    4^{2\gamma}2\pi n_F^2 v_F^2 (az)^{4\gamma}
  }{
    (1+2\gamma)u^2
  }
  e^{i V d/u}
  V
  e^{-4\gamma zd}
  \,.
\end{align}
\end{subequations}
Here $\gamma=\frac{1}{4}(g+g^{-1})-\frac{1}{2}$, $u=v_F/g$ is the
plasmon velocity, and $n_F = L/(2\pi v_F)$ the density of states at
the Fermi energy, and $a$ the short-distance cutoff length.

The finite exponent $4\gamma z d$ occurs due to charge
fractionalization \cite{lehur2005-dom,*lehur2006-eli}. At low temperatures,
$V\ll{}T\ll{}u/d$, the exponential dependence crosses over to a power
law in $T$, as shown in Fig.~\ref{fig:GAB-vs-T}, and $Z_1$ and $Z_2$
coincide.

Note that for $I_{\rm AB}$ , the oscillations dephase exponentially
even in the absence of interactions ($\gamma=0$). The difference
arises because spin-conserving tunneling in a HLL fixes the
direction of propagation, and so the pair of trajectories contributing
to interference has an unavoidable dynamical phase difference of
$2kd$ which also leads to a dependence on $k_F$ \cite{chu2009-coa}
[$I_{\rm AB}\propto\cos(2k_Fd+\varphi_0)$] \footnote{For the sake of clarity, we assume that effective phase factors $e^{i\alpha k_{F} x}$ are the same on both edges, which could be achieved by gate voltages.}, similarly to what
happens in chiral Luttinger liquids \cite{geller1997-aei}.

\begin{figure}
  \includegraphics{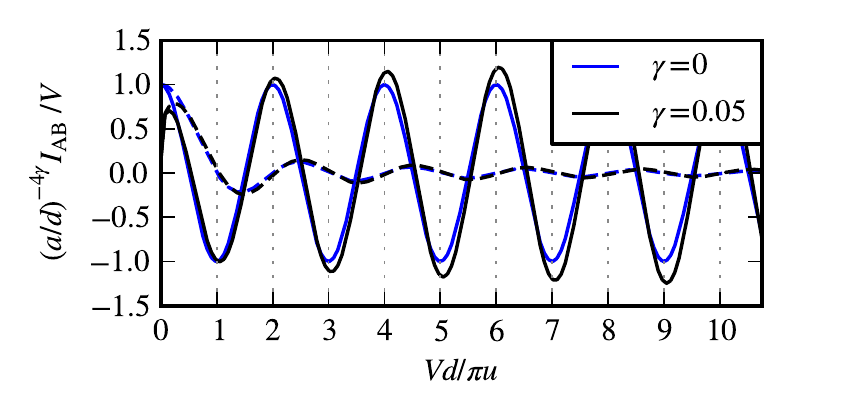}
  \caption{
    \label{fig:GAB-vs-V}
    (Color online)
    Bias dependence of the amplitude of the interference component of
    the current, for $T=0$. The spin-flip (solid),
    non-spin-flip (dashed) contributions are shown. Apart from the
    low-bias power law, the noninteracting ($\gamma=0$, $g=1$) and
    interacting ($\gamma=0.05$, $g\approx0.64$) cases differ mainly in
    the oscillation period $\pi u/d=\pi v_F/gd$.
  }
\end{figure}

Moreover, the interference contributions in the current oscillate not
only with the flux, but also with the bias $V$, as shown in
Fig.~\ref{fig:GAB-vs-V}.  The single characteristic frequency
$\pi{}u/d$ is different from interference in a SFLL, where two
characteristic frequencies exist \cite{peca2003-fia, recher2006-tlc}
because of spin-charge separation.

{\it Relation between charge and spin currents.}  To understand the
dephasing in the spin current, it is useful to note first that in a
HLL the spin current is closely associated with the charge
current. The changes in charge ($c$) and spin ($s$) currents due to
tunneling are
$
\delta I_c \propto
\delta\rho_+ - \delta\rho_-
\,,
$
and
$
\delta I_s \propto
\delta\rho_+ + \delta\rho_-
$
in terms of changes in the densities $\rho$.
Note that this spin current is defined with respect to Kramers partners.
When evaluated in the
noninteracting right (left) lead, in which the $+$ and $-$ modes are
independent and $\delta\rho_-=0$ ($\delta\rho_+=0$) because of
ballistic transport, we find $ \delta I_c^{1R} = \delta I_s^{1R}$
and $ \delta I_c^{1L} = -\delta I_s^{1L} $ on the upper edge~1
(positive direction of current is from left to right in
Fig.~\ref{fig:interferometer}).  On the lower edge~2, the sign of the
spin current is flipped. In a four-terminal setup, this relation
becomes more transparent if one considers an XYZ decomposition
\cite{teo2009-cbo} of the currents, extended to account for
possible non-conservation of spin:
$I^X=\frac{1}{2}[I^{1L}+I^{2L}+I^{1R}+I^{2R}]$
indicates the total current flowing from the left (L) to the right (R),
$I^Y=\frac{1}{2}[I^{1L}-I^{1R}-I^{2L}+I^{2R}]$
the total current from top to bottom,
$I^Z=\frac{1}{2}[I^{1L}+I^{1R}-I^{2L}-I^{2R}]$
the current flowing between the diagonals,
and $I^S=-I^{1L}-I^{2L}+I^{1R}+I^{2R}$ the non-conserving ``source'' current
flowing out of the system. In
this representation we find:
\begin{align}
  \label{eq:spincurrent-xyz}
  \begin{pmatrix}
    \delta I_s^X \\
    \delta I_s^Y \\
    \delta I_s^Z \\
    \delta I_s^S
  \end{pmatrix}
  =
  \begin{pmatrix}
    0 & -1 & 0 \\
    -1 & 0 & 0 \\
    0 & 0 & 0 \\
    0 & 0 & 2
  \end{pmatrix}
  \begin{pmatrix}
    \delta I_c^X \\
    \delta I_c^Y \\
    \delta I_c^Z
  \end{pmatrix}
  \,,
\end{align}
since the dc charge current is conserved ($I_c^S=0$). The result
applies to any helical four-terminal setup. Note that the spin current
flows perpendicular to the charge current, which is a signature of the
spin-Hall effect giving rise to the helical state.  As a result, the
spin current can be accessed by a measurement of the transverse charge
current.

{\it Spin tunneling current.}  From Eq.~\eqref{eq:spincurrent-xyz} it
immediately follows that also spin currents suffer from $e$-$e$
interaction induced dephasing. For instance, consider the spin
tunneling current (Y) that can be generated by applying a bias $V_s$
between left and right (X), $\mu_{1+}=\mu_{2+}=V_s/2$,
$\mu_{1-}=\mu_{2-}=-V_s/2$.  In this configuration, there is no
(Y)-charge current \cite{strom2009-tbe}, similarly as in the (Y)-bias
configuration where there was no spin (Y)-current. Here, the spin
current is given by the difference between particle currents tunneling
into the $+$ and $-$ channels, and it can
be directly evaluated using Eq.~\eqref{eq:channel-current-components}:
$ \avg{\hat{I}_{T,s}} = \avg{-\hat{I}_{+} + \hat{I}_{-}} = I_s^{\rm
  direct} + I_s^{\rm AB}$. To leading order in $H_{T}$,
the spin (Y)-tunneling current has the same form as
the charge (Y)-tunneling current in the Y-biasing configuration (see
Eq.~(\ref{eq:high-t-dephasing-z2})), except that in this order in
$H_T$, spin-flipping tunneling cannot contribute \footnote{ This
  results from the fact that the biasing is symmetric with respect to
  up- and down spins, and so $I_s^S$ vanishes.}.

\begin{figure}
  \includegraphics{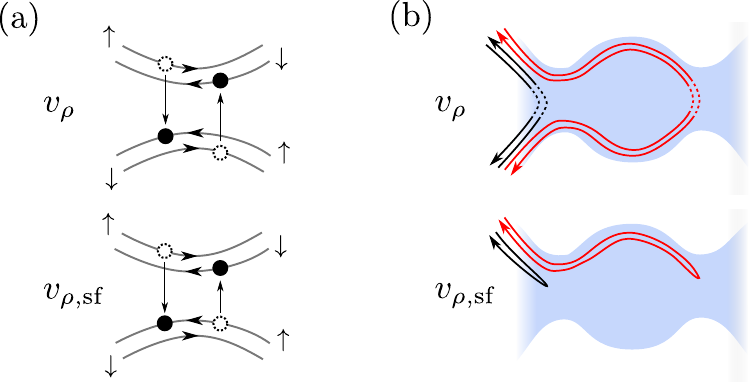}
  \caption{
    \label{fig:tunneling-events}
    (Color online)
    Dominant (neutral) compound tunneling contributions to spin
    interference (Y)-current in (X)-bias configuration.
    (a) Illustrations of compound
    processes. The process $v_{\rho}$ conserves spin, whereas process
    $v_{\rho,{\rm sf}}$ flips one spin. (b) The dynamical phase
    differences of interfering paths for these compound terms with
    arrows describing the propagation directions of electrons [filled
    dots in (a)].
  }
\end{figure}

Depending on the interaction parameter $g$ and the amount of spin-flips in
tunneling, the interference contribution in spin current can at high
temperatures be dominated by {\it two-particle} tunneling processes
(see Fig.~\ref{fig:tunneling-events}), as described by the effective
Hamiltonian (cf. \cite{teo2009-cbo}),
\begin{align}
\label{eq:compound1}
  H_{T}^{(2)} &=
  v_\rho \psi_{1+} \psi_{2-}^\dagger \psi_{2+} \psi_{1-}^\dagger
  \\\notag
  &
  +
  v_{\rho,sf} \sum_{i\ne{}j}\psi_{i,+}\psi_{i,-}^\dagger [
    \psi_{j,+}^\dagger\psi_{j,+}-\psi_{j,-}^\dagger\psi_{j,-}]
  + \mathrm{h.c.}
  \,.
\end{align}
The spin-density fluctuation assisted backscattering process
$v_{\rho,sf}$, which is present only when spin flips are allowed, has
a slower dephasing than single particle processes in the whole range
$1/2<g<1$: $I_s^{\rho, sf, \rm AB}\propto
V_s\cos(V_sd/u)\cos(2k_Fd+\varphi_0)\exp(-2gzd)$. The spin-conserving
process $v_\rho$ on the other hand dominates the single-particle one only
for $g<1/\sqrt{3}$: $I_s^{\rho, \rm AB}\propto
\sin(2V_sd/u)\cos(4k_Fd+\varphi_0)\exp(-4gzd)$.  Although these
processes do not transport charge between the two edges nor couple to
the flux $\Phi$, the oscillation with the bias $V_s$ and $k_{F}$
remains.

Note that the above implies that in a four-terminal HLL setup, charge
and spin tunneling currents have different dephasing exponents, as the
interference contribution is dominantly carried by different types of
excitations: Neutral electron-hole excitations
[Eq.~\eqref{eq:compound1}] for $g<1$ have different correlation
lengths compared to single-particle excitations. Consider, for
instance, the single-particle dephasing exponent $[1+(\frac{g+g^{-1}}{2}-1)]2zd$
from Eq.~\eqref{eq:high-t-dephasing-z2}
compared to the $v_{\rho}$ process, $4gzd=[1 - (1-g)]4zd$, where $2zd$
and $4zd$ arise from dynamical phase differences
[cf. Fig.~\ref{fig:tunneling-events}(b)]. Interestingly, for $v_\rho$
the repulsive interaction {\it reduces} the dephasing from the
noninteracting value.


The above should be contrasted to what occurs in usual SFLLs.  There,
cotunneling of electrons can create an uncharged spin excitation
(spinon), which does not couple to interactions.  The lowest-order
tunneling process coupling only to the spin sector
is the two-electron process
\begin{align}
 \label{eq:effective-spin-tunneling}
  H_{T}^{(2s)} =
  v_\sigma L^2 \sum_{\sigma=\uparrow,\downarrow,\alpha}\psi_{1,\alpha,\sigma}^\dagger\psi_{2,\alpha,\sigma}\psi_{2,\alpha,-\sigma}^\dagger\psi_{1,\alpha,-\sigma} + \mathrm{h.c.}
  \,,
\end{align}
where opposite spins tunnel to opposite edges. A calculation along the
same lines as above yields the interference component in the spin
tunneling current,
\begin{equation}
  \label{eq:spincurrent}
  I_{s,\mathrm{osc}}
  = 4\pi n_F^4|v_\sigma|^2 V_s' [(V_s')^2 + (2\pi T)^2]
  \cos(V_s' d/v_F)
  \,,
\end{equation}
where $V_s' =
\mu_{1\uparrow}-\mu_{1\downarrow}-\mu_{2\uparrow}+\mu_{2\downarrow}$.
Exponential dephasing is indeed absent, as expected under SU(2) spin
symmetry.
%
%
%
%
Generating such spin currents requires a ``spin bias'' $V_s'$, which in our
model is essentially equivalent to a difference in spin densities
between the two edges. In SFLLs one possibility for inducing this is
to contact the SFLL to a system in which a spin imbalance is
externally maintained \cite{zutic2004-sfa}, or to couple it to a HLL
\cite{liu2010-cdi} which may also allow measuring the spin currents
via charge currents.

%
%

In summary, we showed how electron-electron interaction results in
dephasing of interference oscillations in 1D helical liquids, both in
charge and spin tunneling currents, with respective exponents that
can differ. Moreover, we pointed out how the close coupling of the
spin current to the charge current in a helical liquid can result in a
qualitatively different behavior from spinful Luttinger liquids.  Such
effects provide a clear signature of the helicity of the transport, and
understanding them may be valuable for applications.

We thank B. Trauzettel and T. Ojanen for useful discussions, and
acknowledge financial support from the Emmy-Noether program of the
DFG.

\bibliography{helicaledge}

\appendix

\section{Current operators}

In this work we are interested only in the time-averaged spin or
charge currents entering the noninteracting leads.  To compute them
conveniently, we need to identify the operators corresponding to the
tunneling currents, taking into account the plasmon reflections at the
edges of the leads.

The identification can be done using a similar approach as in
Ref.~\cite{safi1995-tts}.  From the Heisenberg equation of motion for
the field $\phi$, under a Hamiltonian $H=H_0+V[\phi,\vartheta]$,
where the effective tunneling term $V$ has support only in a finite
region, one finds an exact result for the change in the charge current
operator $\delta{\hat I}_{c}=
\pi^{-1/2}v_F\partial_x(\phi-\phi\rvert_{V=0})$ in the Heisenberg
picture:
\begin{align}
  \label{eq:plasmon-operator-transmission}
  \delta \hat{I}_{c}(x,t)
  &=
  v_F\int_{-\infty}^\infty\dd{t}'\dd{x}'\sum_{\alpha=\pm}\alpha
  \biggl(
  \frac{1 + g}{2g}D_\alpha(xt,x't')
  \\\notag
  &\;
  - \frac{1-g}{2g}D_{-\alpha}(xt,x't')
  \biggr)\hat{j}_\alpha(x't')
  \,,
  \\
  \label{eq:current-operator-definition}
  \hat{j}_\alpha(x') &= \frac{1}{2\sqrt{\pi}}\Bigl(
  \frac{\delta V}{\delta\phi(x')} + \alpha \frac{\delta V}{\delta\vartheta(x')}
  \Bigr)
  \,,
\end{align}
where $D_\pm$ are solutions to the plasmon wave equation $\partial_t^2
D - \partial_x(v_F^2 g(x)^{-2}\partial_x D) = 0$ with initial conditions
corresponding to right (left) moving $\delta$-pulses starting at
$x=x'$: $D_\pm(x,t',x',t')=\delta(x-x')$,
$\partial_tD_\pm(x,t',x',t')=\mp{}u\delta'(x-x')$.  Reflections at the lead
edges where $g(x)$ changes split the initial pulse to a train of
pulses escaping to the leads. Solving the wave equation, one can
verify that the fraction of an initially right-going pulse entering
the right (left) lead in the long-time limit is
$w_{R+}=\lim_{x\to\infty}\int_{0}^\infty\dd{(t-t')}D_{+}(x,t,x',t')=(1+g)/2$
[$w_{L+}=(1-g)/2$].  Because
\begin{gather}
  \frac{1+g}{2g}\frac{1+g}{2} - \frac{1-g}{2g}\frac{1-g}{2} = 1
  \,,
  \\
  \frac{1+g}{2g}\frac{1-g}{2} - \frac{1-g}{2g}\frac{1+g}{2} = 0
  \,,
\end{gather}
Eq.~\eqref{eq:plasmon-operator-transmission} leads to the conclusion that
(i) all current injected in the $+$ ($-$) channel
enters in the right (left) lead, in the time average, and, (ii)
$\hat{I}_{\alpha}=\int\dd{x'}\hat{j}_\alpha(x')$ is the operator
corresponding to the total current injected into channel~$\alpha$.

Several other points can also be directly read off
\eqref{eq:plasmon-operator-transmission}: since the fluctuation of
spin density in a helical liquid is proportional to the same operator
$\delta\rho_+ - \delta\rho_-$ as the charge current, it can be seen
that a spin injected to a helical liquid fractionalizes into
counterpropagating plasmons carrying the fractional spin
$(g\pm{}1)/2g$ (cf. \cite{das2010-sps}). A similar calculation for the
charge density fluctuation implies that an injected electron
fractionalizes to plasmons carrying the charge $(1\pm{}g)/2$, as usual
\cite{pham2000-fei}. The equation also guarantees that similarly as
charge, a spin injected to the $+$ ($-$) channel will, in the
long-time limit, be transmitted in entirety to the right (left) lead.
This implies that spin fractionalization cannot be detected by simple
time-averaging spin current measurements (such as those proposed
e.g. in \cite{das2010-sps}), but instead one needs to have access to
time scales of $gl/v_F$ characteristic of the plasmon transport.

\section{Scaling dimensions and dephasing}


The limiting behavior of dephasing of interference effects at low or
high temperatures can be found from an analysis of the scaling
dimensions of the terms in the effective tunneling Hamiltonian.

Given a bosonized operator
$M=v e^{in\sqrt{\pi}(\phi+\vartheta)}e^{iq\sqrt{\pi}(\phi-\vartheta)}$
(on a single edge), within our model one finds the correlation
function
\begin{gather}
  \label{eq:scaling-dimension}
  \avg{M(x,t)M(0,0)^\dagger}_0\propto{}
  |v|^2F(x,t)^{2\Delta_+}(F(x,-t)^*)^{2\Delta_-}
  \,,
  \\
  \Delta_+(n,q) = \frac{[(1-g)n + (1+g)q]^2}{8g}
  =
  \Delta_-(q,n)
  \,,
\end{gather}
where $F(x,t)=i a z /\sinh[z(x-ut-ia)]$, $z=\pi T/u$, and $a$ is the
short-distance cutoff length.  The scaling dimension of
operator $M$ is $\Delta = \Delta_+ + \Delta_-$.  Now, if $M$
represents an effective tunneling term in the Hamiltonian, a similar
Kubo calculation as done below indicates that at low temperatures its
contribution to spin/charge conductance scales as $G\propto
|v|^2T^{2\Delta-2}$, but at high temperatures $Td/u\gg1$ interference effects
dephase as $G\propto{}e^{-4\pi T d \Delta_d}$, where the
dephasing exponent $\Delta_d=\min[\Delta_+, \Delta_-]$ can differ from
$\Delta/2$.

Comparing $\Delta_d$ between effective tunneling processes allows us
to identify those dominating at high temperatures, and results
for the scaling dimension $\Delta$ indicate the magnitude of the
tunneling element, and justify the use of perturbation theory.

For $1/2<g<2$ we have $\Delta\ge1$ for all time-reversal symmetric
processes, so that in view of renormalization group flow, the
situation is perturbatively stable, and tunneling scales to zero at
low energies \cite{teo2009-cbo,hou2009-cja}. Concerning dephasing of
interference effects, of the tunneling processes transporting charge,
we find that at high temperatures the single-particle process
dominates (for $g>1/2$),
\begin{align}
  \label{eq:plus-plus-scaling}
  \psi_{1+}\psi_{2+}^\dagger
  \,,
  \qquad
  \Delta_d = (g + g^{-1})/4 - 1/2
  \,.
\end{align}
If spin flips are not allowed, the dominant process is still a
single-particle one
\begin{align}
  \label{eq:plus-minus-scaling}
  \psi_{1+}\psi_{2-}^\dagger
  \,,
  \qquad
  \Delta_d = (g + g^{-1})/4
  \,,
\end{align}
and the situation stays the same in the whole range $1/2<g<2$.

For spin transport, the situation is considerably different: suppose
first that spin is conserved. Then, for $1/\sqrt{3}<g<\sqrt{3}$
single-particle tunneling dominates,
\begin{align}
  \psi_{1+}\psi_{2-}^\dagger
  \,,
  \qquad
  \Delta_d = (g + g^{-1})/4
  \,,
\end{align}
but for $1/2<g<1/\sqrt{3}$ a two-particle cotunneling process has the
lowest exponent,
\begin{align}
  \psi_{1+}\psi_{2-}^\dagger\psi_{2+}\psi_{1-}^\dagger
  \,,
  \qquad
  \Delta_d = g
  \,.
\end{align}
In this case, it occurs that $\Delta_d=\Delta/2$.

If spin flips are allowed, one finds that the process with the smallest
$\Delta_d$ in the whole range $1/2<g<1$ is in fact a two-particle one,
\begin{align}
  \psi_{1+}\psi_{2+}^\dagger\psi_{2+}\psi_{1-}^\dagger
  \,,
  \qquad
  \Delta_d = g/2
  \,.
\end{align}
Note that (i) this process is not important for the charge tunneling current,
since it does not transport charge, and (ii) it is higher order in
tunneling and has a larger scaling dimension $\Delta=1+g$ than the
single particle process. Note that it can also be written in the
explicitly time reversal symmetric form
$\psi_{1+}\psi_{1-}^\dagger[\psi_{2+}^\dagger\psi_{2+}-\psi_{2-}^\dagger\psi_{2-}]+\mathrm{h.c.}$
resembling spin density fluctuation assisted backscattering.

From the above discussion, we conclude that the leading results for
charge tunneling (Y) current are obtained with first-order
perturbation theory, but for spin current also second-order
contributions, or alternatively, the effective 2-particle tunneling,
needs to be analyzed.

\vspace{6ex}

\section{Kubo correlators}


The correlation functions $Z_1$, $Z_2$ appearing in the Kubo
calculation are
\begin{align}
  Z_1(x,\omega) &=
  iL^2
  \int_{-\infty}^\infty\dd{t}e^{i\omega t}\bigl\langle
  [\psi_{1,+}^\dagger(x,t)\psi_{2,+}(x,t),
  \\\notag&\qquad
  \psi_{2,+}^\dagger(0,0)\psi_{1,+}(0,0)]_-
  \bigr\rangle_0
  \,,
  \\
  Z_2(x,\omega) &=
  iL^2e^{2ik_Fx}
  \int_{-\infty}^\infty\dd{t}e^{i\omega t}\bigl\langle
    [\psi_{1,+}^\dagger(x,t)\psi_{2,-}(x,t),
  \\\notag&\qquad
    \psi_{2,-}^\dagger(0,0)\psi_{1,+}(0,0)]_-
    \bigr\rangle_0
    \,.
\end{align}
These correlators can be evaluated via standard bosonization
techniques for spinless Luttinger liquids \cite{giamarchi2004-qpi}
\begin{widetext}
\begin{subequations}
\label{eq:currentcorrelator}
\begin{align}
  Z_1(x,\omega)
  &=
  2i
  n_F^2  v_F^2  a^{-2}
  \int_{-\infty}^\infty\dd{t}
  e^{i\omega t}\Im\biggl[
    \Bigl(
    \frac{
      i a z
    }{
      \sinh\bigl(
      z (x - u t + i a)
      \bigr)
    }
    \Bigr)^{2\gamma+2}
    \Bigl(
    \frac{
      -i a z
    }{
      \sinh\bigl(
      z (x + u t - i a)
      \bigr)
    }
    \Bigr)^{2\gamma}
  \biggr]
  \,,
  \\
  Z_2(x,\omega)
  &=
  2i
  n_F^2  v_F^2  a^{-2}
  \int_{-\infty}^\infty\dd{t}
  e^{i\omega t}\Im\biggl[
    \Bigl(
    \frac{
      i a z
    }{
      \sinh\bigl(
      z (x - u t + i a)
      \bigr)
    }
    \Bigr)^{2\gamma+1}
    \Bigl(
    \frac{
      -i a z
    }{
      \sinh\bigl(
      z (x + u t - i a)
      \bigr)
    }
    \Bigr)^{2\gamma+1}
  \biggr]
  \,.
\end{align}
\end{subequations}
\end{widetext}
Here, $a$ is the short-distance cutoff,
$\gamma=\frac{1}{4}(g+g^{-1})-\frac{1}{2}$, $u=v_F/g$ is the plasmon
velocity, $z=\pi T/u$ the inverse thermal length, and $n_F = L/(2\pi
v_F)$ the density of states at the Fermi energy.

Integrals of the above type can be evaluated in closed form by making
use of the binomial series
\begin{align}
  \label{eq:sinh-binomial-series}
  \left(
    \frac{i}{\sinh \xi}
  \right)^{\alpha}
  =
  2^\alpha
  i^{\alpha(\sgn \Re \xi)}
  \sum_{n=0}^\infty
  \frac{\Gamma(\alpha + n)}{n!\Gamma(\alpha)}e^{-(\alpha + 2n)(\sgn\Re \xi) \xi}
  \,,
\end{align}
and its Fourier transform. In the limit $a\equiv\Im\xi\to0^+$ one finds

\begin{align}
  \label{eq:fraction-ft}
  &
  \int_{-\infty}^\infty
  \dd{t}
  e^{i\omega t}
  \left(
    \frac{i}{\sinh(i a - t)}
  \right)^\alpha
  \\
  \notag
  &
  \simeq
  \frac{
    \frac{
      2^\alpha \pi^2
    }{
      \Gamma(\alpha)
    }
     e^{\pi\omega/2}
  }{
    \Gamma(1+\frac{i\omega}{2}-\frac{\alpha}{2})
    \Gamma(1-\frac{i\omega}{2}-\frac{\alpha}{2})
    [\cosh(\pi\omega)-\cos(\pi\alpha)]
  }
  \,,
\end{align}
which has simple poles at $\omega=i\alpha + 2in$, $n=0,1,2,\ldots$.
We now rewrite \eqref{eq:currentcorrelator} in the form $f(\omega) -
f(-\omega)^*$, where $f$ is a convolution that can be transformed via
contour integration to a Matsubara sum, which can be evaluated. As a
result, we find
\begin{widetext}
\begin{align}
  \label{eq:scorrelator}
  Z_1(x,\omega)
  &=
  \frac{
    4^{2\gamma+1}
    \pi
    n_F^2
    v_F^2
    z
    (az)^{4\gamma}
  }{
    u
  }
  i
  \Biggl(
  \frac{
    e^{-4\gamma xz}
  }{
    \Gamma(2+2\gamma)
  }
  e^{i\omega x}
  \Gamma(1+2\gamma-\frac{i\omega}{2uz})
  {}_2\tilde{F}_1(2\gamma,1+2\gamma-\frac{i\omega}{2uz},-\frac{i\omega}{2uz},e^{-4xz})
  \\\notag
  &\qquad-
  \frac{
    e^{-4(\gamma+1) xz}
  }{
    \Gamma(2\gamma)
  }
  e^{-i\omega x}
  \Gamma(1+2\gamma+\frac{i\omega}{2uz})
  {}_2\tilde{F}_1(2+2\gamma,1+2\gamma+\frac{i\omega}{2uz},2+\frac{i\omega}{2uz},e^{-4xz})
  \Biggr)
  \,,
  \\
  \Re
  Z_2(x,\omega)
  &=
  \frac{
    4^{2\gamma+1}2\pi
    n_F^2
    v_F^2
    z
    (az)^{4\gamma}
  }{
    u
  }
  \frac{
    e^{-4(\gamma+1/2)xz}
  }{
    \Gamma(1+2\gamma)
  }
  \Im\bigl[
  e^{i\omega x}
  \Gamma(1+2\gamma-\frac{i\omega}{2uz})
  {}_2\tilde{F}_1(1+2\gamma,1+2\gamma-\frac{i\omega}{2uz},1-\frac{i\omega}{2uz},e^{-4xz})
  \bigr]
  \,,
\end{align}
\end{widetext}
where ${}_2\tilde{F}_1(a,b,c,z)={}_2F_1(a,b,c,z)/\Gamma(c)$ is the
regularized hypergeometric function.  For $z\gg{}1/d$ and $V\ll{}T$ we
obtain Eqs.~(5a) and~(5b) of the main text.

\section{Compound tunneling between HLLs}

The spin transport due to the compound tunneling Hamiltonian
\begin{align}
\label{eq:compound1copy}
  H_{T}^{(2)} &=
  v_\rho \psi_{1+} \psi_{2-}^\dagger \psi_{2+} \psi_{1-}^\dagger
  \\\notag
  &
  +
  v_{\rho,sf} \sum_{i\ne{}j}\psi_{i,+}\psi_{i,-}^\dagger [
    \psi_{j,+}^\dagger\psi_{j,+}-\psi_{j,-}^\dagger\psi_{j,-}]
  + \mathrm{h.c.}
  \,.
\end{align}
can be handled similarly as above.  For $H_{T}=v_\rho
\psi_{1+}\psi_{1-}^\dagger \psi_{2+}\psi_{2-}^\dagger+\mathrm{h.c.}$
we find $I_+ = -I_- = 2iv_\rho \psi_{1+}\psi_{1-}^\dagger
\psi_{2+}\psi_{2-}^\dagger+\mathrm{h.c.}$ on the basis of
Eq.~\eqref{eq:current-operator-definition}.  From this we immediately
see that in the X-bias configuration the charge (Y) current vanishes,
and the spin (Y) current is obtained by a straightforward calculation:
\begin{align}
  \avg{I_{T,s}}
  &=
  2 \sum_{R=0,d}|v_\rho(R)|^2\Re{} Z_3(0, -2V_s)
  \\
  \notag
  &\quad+ 4\Re[v_\rho(0)v_\rho(d)^*e^{4ik_Fd}] \Re{} Z_3(d, -2V_s)
  \,,
  \\
  Z_3(x,\omega) &= (2\pi a)^{-2}Z_2(x,\omega)\rvert_{\gamma\mapsto{}g-1/2}
  \,.
\end{align}
Apart from the prefactor, the result is identical to $Z_2$ with a
different exponent, in agreement with
Eq.~\eqref{eq:scaling-dimension}.

Similarly, starting from
\begin{align}
  H_{T}^{(2)} =
  \sum_{i\ne{}j}v_{\rho,sf}\psi_{i,+}\psi_{i,-}^\dagger [
  \psi_{j,+}^\dagger\psi_{j,+}-\psi_{j,-}^\dagger\psi_{j,-}]
  + \mathrm{h.c.}
  \,,
\end{align}
we get $I_{1+}=-I_{1-}=2iv_{\rho,sf} \psi_{1+}\psi_{1-}^\dagger
[\psi_{2+}^\dagger\psi_{2+}-\psi_{2-}^\dagger\psi_{2-}]+\mathrm{h.c.}
\equiv{}iM_1-iM_1^\dagger$, and similarly on edge 2. The relevant
correlation functions are of the form
\begin{align}
  &\avg{M_1(r,t) M_1^\dagger(0,0)}_0
  \propto
  |v_{\rho,\rm sf}|^2
  e^{-i(\mu_{1+}-\mu_{1-})t}e^{2ik_Fr}
  \\\notag
  &\;\times
  F(r,t)^{g}(F(r,-t)^*)^{g}[F(r,t)^2+(F(r,-t)^*)^2]
  \,,
\end{align}
which implies that the high-temperature dephasing exponent is
$\Delta_d=g/2$, being different from $\Delta/2=g/2+1$. We remark
that the contribution in third order in tunneling vanishes identically
due to particle conservation.

Note that in the X-bias configuration, we have $\delta I_c^X=2I_+$,
$\delta I_c^Y=0$, and $\delta I_c^Z=0$ for both of the above processes.

\section{Description of SFLL}

In a spinful Luttinger liquid \cite{giamarchi2004-qpi}, the direction
of spin and motion is not coupled and the fermion operator has to be
defined separately for spin and direction of motion (at each edge):
\begin{equation}
  \label{eq:spinfermioperator1}
  \psi_{\alpha,\uparrow(\downarrow)}(x)\simeq e^{i\alpha[k_Fx
    +\sqrt{\pi}\vartheta_{\uparrow(\downarrow)}(x)]+i\sqrt{\pi}\phi_{\uparrow(\downarrow)}(x)}
  \,.
\end{equation}
It is advantageous to change to charge ($\rho$) and spin ($\sigma)$
variables
$\phi_{\uparrow(\downarrow)}=(1/\sqrt{2})(\phi_{\rho}\pm\phi_{\sigma})$
and
$\vartheta_{\uparrow(\downarrow)}=(1/\sqrt{2})(\vartheta_{\rho}\pm\vartheta_{\sigma})$,
explicitly
\begin{equation}
\label{eq:spinfermioperator2}
\psi_{\alpha,\sigma}(x)\simeq e^{i\alpha k_{F}x}e^{i\sqrt{\frac{\pi}{2}}(\phi_{\rho}+\alpha\vartheta_{\rho}+s(\phi_{\sigma}+\alpha\vartheta_{\sigma}))}
\,,
\end{equation}
with $s=\pm$ for $\sigma=\uparrow,\downarrow$.
In these new variables, the interacting system generally splits into a
sum of spin and charge parts: $H_{\rm SFLL}=H_{\rho}+H_{\sigma}+g_1
\int_{-\infty}^{\infty} dx \cos(\sqrt{8\pi}\vartheta_{\sigma})$. These
Hamiltonians are characterized by interaction parameters $g_{\nu}$ and
velocities $v_{\nu}$, $\nu=\rho,\sigma$. At the low-energy fixed
point, $g_{1}$ flows to zero for repulsive SU(2) invariant
interactions and therefore the Hamiltonian becomes $H=H_{\rm
  SLL}=H_{\rho}+H_{\sigma}$ with
\begin{equation}
\label{eq:hcharge}
H_{\rho}=\frac{u_{\rho}}{2}\int_{-\infty}^{\infty} dx \left[g_{\rho}\left(\partial_{x}\phi_{\rho}\right)^2+\frac{1}{g_{\rho}}\left(\partial_{x}\vartheta_{\rho}\right)^{2}\right],
\end{equation}
and
\begin{equation}
\label{eq:hspin}
H_{\sigma}=\frac{u_{\sigma}}{2}\int_{-\infty}^{\infty} dx \left[g_{\sigma}\left(\partial_{x}\phi_{\sigma}\right)^2+\frac{1}{g_{\sigma}}\left(\partial_{x}\vartheta_{\sigma}\right)^{2}\right],
\end{equation}
where $g_{\rho}\equiv g$, $u_{\rho}\equiv u$, and $g_{\sigma}=1$
\cite{giamarchi2004-qpi}.  The spin sector
therefore becomes effectively non-interacting.

The dephasing for spin currents can therefore be absent in second
order in the tunneling which we describe by the effective Hamiltonian
\begin{align}
 \label{eq:effective-spin-tunnelingcopy}
  H_{T}^{(2s)} =
  v_\sigma L^2 \sum_{\sigma=\uparrow,\downarrow,\alpha}\psi_{1,\alpha,\sigma}^\dagger\psi_{2,\alpha,\sigma}\psi_{2,\alpha,-\sigma}^\dagger\psi_{1,\alpha,-\sigma} + \mathrm{h.c.}
  \,,
\end{align}
e.g. consider the term
\begin{equation}
\label{eq:absent}
\psi_{1+\uparrow}\psi_{2+\uparrow}^{\dagger} \psi_{2+\downarrow}\psi_{1+\downarrow}^{\dagger}
\sim e^{i\sqrt{\pi/2}(\phi_{1\sigma}+\vartheta_{1\sigma}-\phi_{2\sigma}-\vartheta_{2\sigma})}.
\end{equation}
The correlation function of terms like this one will not show
fractionalization because $g_{\sigma}=1$ and the charge fields
have disappeared from the bosonized expression.

The coupling constant $v_{\sigma}$ in the effective Hamiltonian
Eq.~\eqref{eq:effective-spin-tunnelingcopy} is second order in the bare
tunneling amplitude and can depend (at low bias voltages) on
temperature in a power-law fashion (see section below) but will not depend on
the interferometer length $d$ as long as $Td/u\gg{}1$.

\section{Renormalization of tunneling operators}

For completeness, we now summarize how compound tunneling terms can
be derived via the real-space perturbative renormalization group
\cite{anderson1970-eri}, which has proved useful in studies of
Luttinger liquids \cite{kane1992-ttb,giamarchi2004-qpi}.

We first note that bosonization allows evaluation of correlation
functions involving Fermi operators in closed form.  In general, for a
set of bosonic operators $\Phi_j$, a well-known expansion applies:
\begin{align}
  \label{eq:bosonic-correlation-expansion}
  \avg{\mathbb{T}\prod_i e^{i \Phi_i}}_0
  =
  e^{-\frac{1}{2}\sum_{ij}\avg{\mathbb{T}\Phi_i \Phi_j}_0}
  \,,
\end{align}
where $\mathbb{T}$ indicates (contour-)time ordering.  Given the bosonic
correlation functions, $ \avg{\Phi_{\pm}(t)\Phi_{\pm}(0)}_0 =
\frac{1}{2\pi{}g} \ln\{ -i z L / \sinh[z (t - i a)] \} $ for
$\Phi_{\pm} \equiv 2^{-1/2}(\phi \mp \vartheta/g)$, one finds a
short-time expansion $\avg{\mathbb{T}A(t)B(t')Q}_0 \propto
|t-t'|^{-\alpha}\avg{\mathbb{T}A(t)B(t)Q}_0$ as $t'\to{}t$ for groups $A$ and
$B$ and $Q$ of Fermi operators.  Since the scaling at $t'\to{}t$ is
independent of $Q$, in view of
Eq.~\eqref{eq:bosonic-correlation-expansion}, this can be understood
as an operator product expansion $A(t)B(t') = |t-t'|^{-\alpha}(AB)(t)$
as $t'\to{}t$.

The perturbative renormalization group now proceeds by integrating out
all short-time divergences at times $a+\dd{a}>|t-t'|>a$ appearing in
the second order of perturbation expansions. The generated compound
terms are absorbed in the first order of the expansion via a change in
the Hamiltonian, $H\mapsto{}H+\dd{H}$, which after rescaling the
cutoff $a\mapsto{}a+\dd{a}$ leads to the RG flow equations.

In this approach, we find the flow equations
corresponding to the effective Hamiltonian
$H_{\rm eff}=H_0+H_T+H_T^{(2)}$,
\begin{align}
  \frac{\dd{t_{\alpha\beta}}}{\dd{\ln a}} &= [1 - (g^{-1}+g)/2] t_{\alpha\beta}
  \,,
  \\
  \frac{\dd{v_\rho}}{\dd{\ln a}} &= [2 - 2g]v_\rho
  - c a t_{+-}t_{-+}^*\theta(g^{-1}-g)
  \,,
  \\
  \frac{\dd{v_{\rho,\rm sf}}}{\dd{\ln a}}
  &= [1 - g] v_{\rho,\rm sf} - c' a t_{++}t_{-+}^*\theta(g^{-1}-1)
  \,,
\end{align}
where the first terms appear, as usual, from the intrinsic cutoff
dependence of the operators, $\avg{\prod_j
  e^{i\Phi_j}}_0\propto{}a^{\Delta}$, and the latter from the short-time
divergences. The factors $c$ and $c'$ are numerical constants.
Note that the effective Hamiltonian is written in terms of Fermi operators.
Moreover, the compound terms above are generated only for repulsive
interactions, $g<1$ --- for $g\ge1$ the perturbation expansion does
not have the corresponding short-time divergences.

The flow equation can be integrated starting from the bare values
$a=a_0$, $t_{\alpha\beta}(a_0)=t_{\alpha\beta}$,
$v_\rho(a_0)=v_{\rho,\rm sf}(a_0)=0$, up to the length scale at
which the operator product expansion breaks down. In our case it is
the size of the interferometer $d$ or the thermal length $1/z$,
whichever is smaller. This yields the
scaling of the prefactors with the cutoff (assuming $1/z\ll d$):
\begin{align}
  t_{\alpha\beta} &\propto (a_0 T)^{(g^{-1}+g)/2-1}t_{\alpha\beta}(0)
  \,,
  \\
  v_\rho &\propto (a_0 T)^{\min(2g-1, g^{-1}+g-2)} T^{-1} t_{+-}(0)t_{-+}^*(0)
  \,,
  \\
  v_{\rho,\rm sf} &\propto (a_0 T)^{\min(g, g^{-1}+g-2)}T^{-1} t_{++}(0)t_{-+}^*(0)
  \,.
\end{align}
A similar treatment gives the scaling for the process described by Eq.~(\ref{eq:effective-spin-tunnelingcopy}) in a SFLL
\begin{align}
  v_\sigma
  &\propto
  (a_0 T)^{\min(1,(g^{-1}+g)/2-1)}
  T^{-1}
  t_{+\uparrow,+\uparrow}(0)t_{+\downarrow,+\downarrow}^*(0)
  \,,
\end{align}
The result for $v_\rho$ coincides with that obtained in
Ref.~\cite{teo2009-cbo} for $g<(\sqrt{5}-1)/2$ at which the scaling
$2g-1$ of the compound process starts to dominate the
scaling $g^{-1}+g-2$ of two single-particle events. The dephasing
exponents for the above processes, on the other hand, follow from the
arguments in the previous sections, and in general they are independent
of the scaling dimensions of the prefactor.

\end{document}